\documentstyle[12pt]{article}

\begin{document}

\author{George Vlasov \\
Landau Institute for Theoretical Physics and \\
Moscow Aviation Institute, \\
Moscow, Russia\thanks{
E-mail: vs@itp.ac.ru}}
\title{Chaos as a basis of new principle for detecting the gravitational
waves}
\maketitle

\begin{abstract}
A particular example of chaos can be conceived in the interaction of
non-linear
oscillator with a harmonic gravitational wave. When we replace the
linear potential
forces by the therm SIN(x), the type of solution becomes subject to
external perturbation.
Although the perturbation produced by the gravitational wave is weak the
standard
estimations allow to predict the appearance of chaos at definite range
of parameters.
This qualitative change in the character
of motion immediately detects the fact of impact with gravitational
wave.
Another advantage relates to a broad range
of frequencies so that the narrow resonance band is not required.
\end{abstract}


\sloppy

The gravitational wave detectors \cite{Weber60,Weber64,Thorne80} are
under
intent achievement for almost thirty years. Laser interferometry \cite
{B+90,A+92,A+96,N+96,Allen96,PGM98} and resonant mass antennas \cite
{BT85,JM93,ZM95,Lobo95,HSP96,MJ97,Stevenson97} are considered as the
most
perspective approaches on the way of discovering gravitational
radiation. As
a matter of fact, resonance in a vibrating system, whichever different
phenomena give rise to vibration, is the basic principle of this
approach.
Thus, a higher sensitivity of the detector, a higher amplification and a

frequency close to that of the gravitational wave, not known beforehand,
are
required. These are the most difficulties which inevitably impede the
researchers. In the present study we propose a principally new method
for
detecting the gravitational waves. Instead of resonance in the linear
oscillator we appeal to a non-linear system vibrating near the
separatrix.
The gravitational wave, then, may set a system to a dynamic chaos
directly
distinguished from the former oscillatory regime. With it, this
qualitative
response occurs even for a considerable (appreciated by orders)
difference
in frequencies of the oscillator and the external agent.

Let a linear polarized gravitational wave propagate along axis $x$ and
collide with detector. As a particular instance which admit a wider
application we consider a pendulum oscillating in $xy$ plane. The
external
force from the gravitational wave with amplitude $2\sqrt{2}A$ is
expressed
through the metric tensor $h_{jk}$ as \cite{Speliotopoulos95}
\[
F_j\left( T\right) =\frac 12\,m\,\ddot h_{jk}\,x^k
\]
and then reduced to
\begin{equation}
F_x\left( T\right) =\,m\,A\,\omega ^2\sin \omega T\ \ l\sin \varphi
\qquad
F_y=0  \label{F}
\end{equation}
for polarization states $\varepsilon _{+}=\varepsilon _{\times
}=1/\sqrt{2}$
and sinusoidal profile of the wave. The equation of motion will look
like
\begin{equation}
ml\ddot \varphi +mg\sin \varphi =F_x\left( T\right)  \label{motion}
\end{equation}
where $\varphi $ is the angular deviation from the equilibrium point,
and $l$
is the length of the pendulum. The suspected values of parameters in
(\ref{F}%
) are \cite{Thorne80}
\begin{equation}
A=10^{-17}\div 10^{-22}\qquad \omega =1\div 10^3\,s^{-1}  \label{A}
\end{equation}

If we introduce the dimensionless time $t=T/\Omega $, where $\Omega
=\sqrt{%
g/l}$, the equations (\ref{motion}) and (\ref{F}) lead to
\begin{equation}
\ddot \varphi +\sin \varphi =A\nu ^2\sin \varphi \sin \nu t\qquad \qquad
\nu
=\frac \omega \Omega  \label{mot2}
\end{equation}
The common principal for detecting the GW is based on resonance in the
linearized system (\ref{mot2}) when $\nu \rightarrow 1$. However, we
shall
consider a dynamic chaos in a nonlinear system. The behavior of system
(\ref
{mot2}) is determined by the integral of motion, namely the initial
energy $%
E_0$ \cite{ZSUC91}. The separatrix value $E_0=1$ is the bound between
two
types of motion: oscillation at $E_0<1$ and rotation at $E_0>1$. As soon
as
the external force acts on the system, the energy will not be conserved.

Nevertheless, a weak perturbation caused by a real gravitational wave
will
not change the character of regular motion (rotation or oscillation) at
a
relatively small $\left( E_0\ll 1\right) $ or very great $\left( E_0\gg
1\right) $ initial energy. On the other hand, even a weak perturbation
becomes sufficient and may lead to transition from rotation to
oscillation
(and controversially) for the motion corresponding to small $|E_0-1|$.
One
cannot predict the behavior of the system, whether it oscillates or
rotates,
for it is determined by the value of energy $E\left( t_n\right) $ at the
$n$%
-th quasiperiod that cannot be known beforehand but must be found step
by
step for each $n$. By the way, the quasiperiod grows monotonously with
the
increase of difference $|E_0-1|$, so that a long quasiperiod hints on a
favorable condition to set the system into a regime of random motion by
a
relatively weak perturbation. Thus, one can say that dynamic chaos
occurs in
the system \cite{ZSUC91}. The system is inclined to chaos if
\begin{equation}
|E\left( t_{n-1}\right) -1|\leq \nu \ |E\left( t_n\right) -E\left(
t_{n-1}\right) |  \label{cr}
\end{equation}
where
\begin{equation}
E\left( t\right) =E_0+A\nu ^2J\left( t\right)  \label{E}
\end{equation}
\begin{equation}
J\left( t\right) =\int\limits_0^t\dot \varphi \left( \tau \right) \sin
\varphi \left( \tau \right) \cos \nu \tau \/\ d\tau  \label{j}
\end{equation}
and the $n$-th quasiperiod is determined by the formula
\begin{equation}
t_n=t_{n-1}+\ln \frac{32}{|E\left( t_{n-1}\right) -1|}  \label{tn}
\end{equation}
while $\varphi \left( t\right) $ is given through the elliptic Jacobi
function:
\begin{equation}
\dot \varphi =\sum_n\frac{2\,\left( -1\right) ^2}{\cosh \left(
t-t_n\right) }
\label{phi}
\end{equation}
Contrary to the usual resonance detectors, the limit $\nu \rightarrow 1$
is
quite insufficient here, since the response of the system on external
perturbation, in the right side of (\ref{mot2}), occurred as transition
to a
chaotic regime, takes place in the sufficiently wide range of external
frequencies. Even if $\omega $ formidably exceeds $\Omega $ or is
negligible
in respect to the latter, the criterion (\ref{cr}) does not require a
great
magnitude of the external force. This property is very useful in view of
the
unknown preset GW frequency being received.

Indeed, a chaos may be set at large $n$, however, if it is assumed to
appear
in the first period, we substitute $n=1$ in Eqs. (\ref{cr})-(\ref{phi}).
The
integral (\ref{j}) can be evaluated analytically in two limit cases,
namely,
\begin{equation}
E_1-E_0=A\nu Q\left( \nu \right) \sin \nu t_1  \label{e1}
\end{equation}
where $Q\left( \nu \right) =4$ for small $\nu $ and
\begin{equation}
Q\left( \nu \right) =\nu ^2\sqrt{\pi }e^2e^{-\pi \nu /2}  \label{Q}
\end{equation}
in the opposite limit. Therefore,
\begin{equation}
|E_0-1|\leq \nu AQ\left( \nu \right) \sin \left( \nu \ln
\frac{32}{|E\left(
t_0\right) -1|}\right)  \label{cr1}
\end{equation}
For example, at $A=10^{-20}$, $\nu =0.1$ and $\nu =10$ we find from
(\ref{tn}%
), (\ref{e1})-(\ref{cr1}) $t_1\simeq 50.5$ and $t_1\simeq 55.8$
respectively. Thus, if the period of the non-perturbed pendulum is
approximately $t_1/P\approx 8$ times as large as the period $P=2\pi $ of
the
linear oscillator, with the same parameters [$\sin \varphi $ is replaced
by
a mere $\varphi $ in Eq. (\ref{motion})], then the interaction of the
gravitational wave with the system will bring dynamic chaos in one
cycle.
For a sufficiently lesser $t_1$ (i.e. relatively large $|E\left(
t_0\right)
-1|$) the chaotic regime does not occur.

Hence, no amplification of the output signal is required, since the
system
will response to GW by transition to a new regime of motion. Although a
high
quality is needed: the energy loss during one quasiperiod must not
exceed
the increment (\ref{e1}). For instance, the thermal noise must not
obscure
the chaos. In other words the energy fluctuation $\langle \Delta
E\rangle $
of the non-linear pendulum must be less than the difference $|E\left(
t_0\right) -1|$. A plain calculation by standard thermodynamic formula
gives
$\langle \Delta E\rangle \simeq 0.5\Theta $, where $\Theta $ is
temperature.
Hence, $\Theta <2mgl|E\left( t_0\right) -1|\sim mglA\sim AE_0$. This
constraint implies $\Theta <10^{-1}\,^{\circ }$K for $A=10^{-20}$,
$m=1$gr, $%
l=10\,$cm and terrestrial value of $g=100\,$cm/s$^2$ .

So, the interaction of GW with the non-linear pendulum, described by the

equation of motion (\ref{motion}), may be directly observed as
transition to
a dynamic chaos if the period of non-perturbed pendulum exceeds the
definite
value determined by Eqs. (\ref{cr})-(\ref{phi}). The frequency of GW is
not
very sufficient and may vary by orders. Of course, the example
considered
above is not a single variant; the results of the present study can be
applied, in principle, to any system where a dynamical chaos is
plausible,
for instance, even when considering an electron in the crystal lattice
\cite
{HM93}.

\end{document}